\documentclass[conference]{IEEEtran}

\IEEEoverridecommandlockouts
\usepackage{pgfplots}
\usepackage{subfigure}
\usepackage{tikz}
\usepackage{textcomp}
\usepackage{cite}
\usepackage{amsmath,amssymb,amsfonts}
\usepackage{algorithm,tcolorbox}
\usepackage{xcolor}
\usepackage{setspace}
\usepackage{url}
\usepackage{float}
\usepackage{algorithmicx}  
\usepackage{algpseudocode}

\definecolor{bggreen}{RGB}{154,205,50}
\definecolor{bgred}{RGB}{255,130,71}
\definecolor{bgbrown}{RGB}{50,205,50}
\usepackage{graphicx}
\usepackage{epstopdf}
\DeclareGraphicsExtensions{.png,.pdf}
\usepackage{amsfonts,latexsym,amssymb,shadow,amsmath,wrapfig,eepicemu}%,psfrag}
\usepackage[margin=0pt,font=small,labelfont=bf,justification=raggedright,format=hang]{caption}
\usepackage[usenames,dvipsnames]{}

\newcommand{\beq}{\begin{equation}}
\newcommand{\eeq}{\end{equation}}
\newcommand{\beqa}{\begin{eqnarray}}
\newcommand{\eeqa}{\end{eqnarray}}
\newcommand{\beqan}{\begin{eqnarray*}}
\newcommand{\eeqan}{\end{eqnarray*}}

\newcounter{l1}
\newcounter{l2}
\newcounter{l3}
\newcommand{\bdotlist}{\begin{list}{$\bullet$}{}}
\newcommand{\bboxlist}{\begin{list}{$\Box$}{}}
\newcommand{\bbboxlist}{\begin{list}{\raisebox{.005in}{{\tiny $\blacksquare$ \ \ }}}{}}
\newcommand{\bdashlist}{\begin{list}{$-$}{} }
\newcommand{\blist}{\begin{list}{}{} }
\newcommand{\barablist}{\begin{list}{\arabic{l1}}{\usecounter{l1}}}
\newcommand{\balphlist}{\begin{list}{(\alph{l2})}{\usecounter{l2}}}
\newcommand{\bAlphlist}{\begin{list}{\Alph{l2}.}{\usecounter{l2}}}
\newcommand{\bdiamlist}{\begin{list}{$\diamond$}{}}
\newcommand{\bromalist}{\begin{list}{(\roman{l3})}{\usecounter{l3}}}

\newtheorem{theorem}{Theorem}[section]
\newtheorem{exercise}[theorem]{Exercise}
\newtheorem{lemma}[theorem]{Lemma}
\newtheorem{proposition}[theorem]{Proposition}
\newtheorem{corollary}[theorem]{Corollary}
\newtheorem{definition}[theorem]{Definition}
\newtheorem{remark}[theorem]{Remark}
\newtheorem{example}[theorem]{Example}

\allowdisplaybreaks[4]
\def\BibTeX{{\rm B\kern-.05em{\sc i\kern-.025em b}\kern-.08em
    T\kern-.1667em\lower.7ex\hbox{E}\kern-.125emX}}
    
\pgfplotsset{compat = 1.14}

\begin{document}

\title{\huge An Algorithmic View on Optimal Storage Sizing}

\author{Zhiqi Wang, Kui Wang, Yang Yu, \emph{and} Chenye Wu \vspace{-0.5cm}
\thanks{The authors are with the Institute for Interdisciplinary Information Sciences (IIIS), Tsinghua University, Beijing, China, 100084. C. Wu is the correspondence author. Email: chenyewu@tsinghua.edu.cn.}
\thanks{This work has been supported in part by National Key R\&D Program of China (2018YFC0809400), a gift fund from Nanjing, and Zhongguancun Haihua Institute for Frontier Information Technology.}}

\maketitle

\begin{abstract}
Users can arbitrage against Time-of-Use (ToU) pricing with storage by charging in off-peak period and discharge in peak periods. In this paper we design the optimal control policy and the solve optimal investment for general ToU scheme. We formulate the problem as dynamic programming for efficient solution. Our result is feasible facing multi-peaked ToU scheme. Simulation studies examine how the user's cost varies with respect to the user's demand randomness; we also demonstrate the performance of our scheme when aggregating users for extra savings.
\end{abstract}

\vspace{-0.2cm}
\section{Introduction}

High renewable penetration warrants a flexible power system, and storage devices provide most flexibility in the future. The renewables also bring significant uncertainties to power system, which implies dynamic pricing is ideal to reflect the real time market conditions. Hence, storage devices are both desired for future power system as well as future electricity markets.
\vspace{-0.2cm}
\subsection{Opportunities and Challenges}

In this paper, we take the first step towards understanding the optimal arbitrage policy with storage system against dynamic pricing. We simplify the problem by eliminating the randomness in dynamic pricing. More precisely, we focus on arbitraging against the general Time-of-Use (ToU) pricing schemes.

Such arbitrage opportunities have been found all over the world (e.g., in US\cite{walawalkar2007economics}, EU\cite{zafirakis2016value}, and China\cite{lin2017economic}). However, to design the arbitrage control policy is challenging since the decisions at each period of the ToU scheme are coupled together. When the ToU scheme consists of multiple peaks, the task is even more delicate.

We formulate the problem as dynamic programming (DP) and propose an efficient arbitrage policy for utilizing the storage system. Based on this arbitrage policy, we introduce a binary search algorithm to decide optimal investment storage capacity (storage sizing) in terms of minimizing electricity bills.

\subsection{Related Work} 
\vspace{-0.2cm}
The major body of literature is on designing the storage control policies under various pricing schemes. Just to name a few, Koutsopoulos \emph{et al.} propose a control policy for finite capacity storage system for utilities \cite{6102369}. Van de ven \emph{et al.} introduce a threshold-based optimal storage control policy facing Markovian prices and demands\cite{6477197}. Qin \emph{et al.} design a sub-optimal online greedy algorithm for dynamic pricing and assuming limited information on the demand \cite{qin2016online}. Wu \emph{et al.} solve the optimal storage sizing problem in 3-tier ToU pricing \cite{3tiercontrol}. Different from previous work, we focus on designing the arbitrage policy and solving the optimizing sizing problem in \emph{general (multi-peaked)} ToU schemes.

%Researchers have designed control policies for different pricing schemes. Wu and Yu propose an optimal policy to arbitrage against 3-tier ToU pricing \cite{3tiercontrol}. For dynamic pricing, Qin \emph{et al.} propose an online modified greedy algorithm and prove its sub-optimality compared to offline \cite{qin2016online}. Van de ven \emph{et al.} propose an optimal control policy for storage charging and discharging under Markovian random rates and demands \cite{6477197}. To the best of our knowledge, optimal control policy for general ToU has not been fully investigated.

Next, we introduce system model in Section \ref{sec:systemmodel}. Section \ref{sec:mainresult} designs the optimal control policy and derives optimal investment in general ToU pricing. Simulation studies are conducted in Section \ref{sec:simulation}. Section \ref{sec:conclusion} provides the concluding remarks. Due to space limitations, we provide the sketch of all the proofs in Appendix, and the full proofs in online supplementary material \cite{S_M}.
\vspace{-0.2cm}
\section{System Model}
\label{sec:systemmodel}
%For simple 2-tier ToU pricing, it suffices to consider the demand of one peak period thus a greedy algorithm can achieve optimal. However, we need to consider the demands and decisions for more periods when facing multi-tier ToU pricing.

%\subsection{Basic Model}

Consider a multi-peaked ToU scheme, where each day is divided into $n$ periods and the electricity rate in $i^{th}$ period is denoted by $\pi_i$. We assume each day ends with the off-peak period, i.e., $\pi_n=\min_i\{\pi_i\}$.

Facing such a ToU scheme, a user consumes random demand $X_i$ in $i^{th}$ period. Let $f_i(\cdot)$ be the probability density function \emph{(pdf)} of $X_i$. Based on the ToU prices and demand distributions, the user may invest a storage with capacity $C$ to reduce its electricity cost. To better utilize the storage, the user need decide how much to charge or discharge at each period. Such decision problems are coupled together through the capacity constraint, i.e., the energy in the storage system cannot be greater than $C$ at all times.

%The cost for the storage is $\pi_s$/kWh per day amortized over its lifespan which leads to the total cost for the storage is $\pi_sC$ per day.

%However, it  The aim is to satisfy the demands with lowest cost under this constraint.

\subsection{Assumptions}

In order to highlight the essence of the problem and simplify our analysis, we make the following assumptions.

\textbf{A1}. The demand is inelastic.

\textbf{A2}. Storage system is lossless, and perfectly efficient in charging and discharging.

\textbf{A3}. The \emph{pdf} of random demand,  $f_i(\cdot)$, is continuously differentiable, and we assume $f_i(x)>0\Leftrightarrow x\geq 0$.

\textbf{A4}. The demands for different periods are independent.

The first three assumptions are standard in the literature. As for the last one, we will examine the case when the demands are dependent across periods in the simulation.

\begin{figure}
    \centering
    \includegraphics[width=6.4cm]{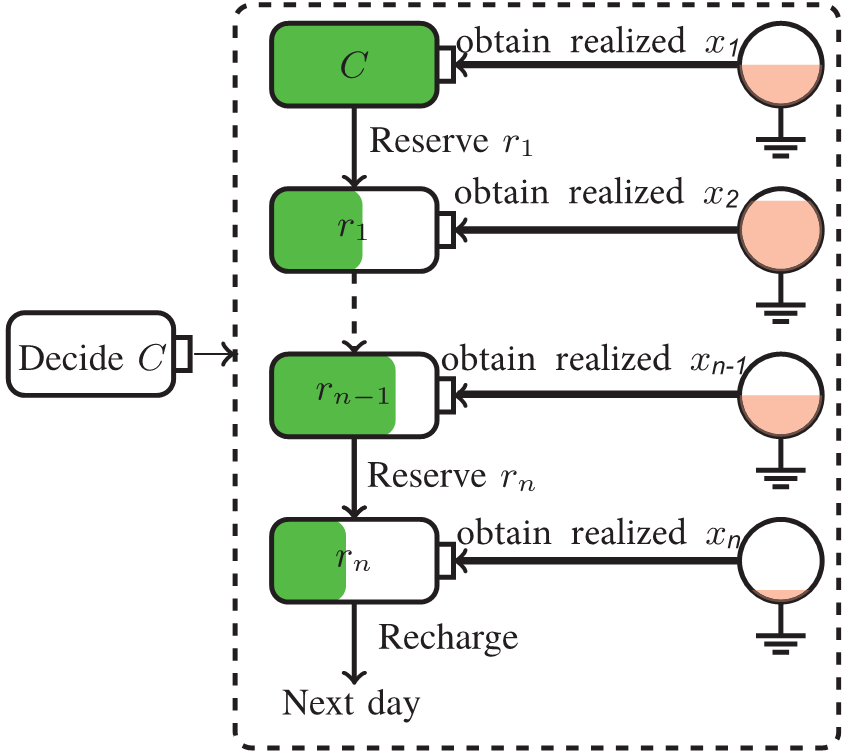}
    \caption{Decision process and DP formulation.}\vspace{-0.4cm}
    \label{fig:my_label}
\end{figure}

%\vspace{-0.3cm}
\subsection{Dynamic Programming Formulation}

We consider two sets of decisions: one is the optimal investment for storage, and the other one is the optimal control policy across multiple periods. As for the system model, we focus on understanding the second task, which serves as the basis for solving the first one. 

To solve a decision problem across multiple periods, one powerful technique is DP, which simplifies a complicated problem by breaking it down into simpler sub-problems in a \emph{recursive} manner. In our model, the decision problem is for each user to minimize the \emph{expected} total cost, while the control space consists of the storage operation strategies at each time slot.

In this paper, we employ a special form of storage operation: reserving energy for future use. That is, at $i^{th}$ period, the user need decide its reservation $r_i$ for future use based on the reservation of $(i-1)^{th}$ period (i.e., $r_{i-1}$), and the random demand $X_i$. We assume the storage is fully charged at the beginning of each day since each day ends with an off-peak period, and the user will always fully charge the battery during the off-peak. We choose to use $r_i's$ to describe the states in DP. It is straightforward to see that the boundary conditions $r_0$ and $r_n$ are both $C$.

%We define the state of $i^{th}$ period by a tuple $(r_{i-1},x_i)$ where $r_i$ represents the reservation of ${(i-1)}^{th}$ period and $x_i$ is the realized demand of $X_i$. We identify the control at $i^{th}$ period is to reserve $r_{i}$ unit energy. We assume the storage is fully charged at the beginning of each day, i.e., the initial state  $(r_0,x_1)$ is $(C,x_1)$.

%The state of this dynamic programming evolves as the random demand at each period $X_i$ becomes realized by $x_i$. The next state consists of the decision of this period and the realized demand of next period $(r_{i+1},x_{i+1})$.  
% This leads to the transition function:
% \begin{align}
%     (r_{i+1},x_{i+1}) = (r_i-x_i+u_i,x_{i+1}).
% \end{align}

Next, we need identify the cost function for DP: the cost at $i^{th}$ period is due to purchasing $r_i+X_i-r_{i-1}$ amount of energy from the main grid. Hence, the \emph{expected} total cost for each user is as follows:
%In our model, the cost in the first period is $\pi_s C$, and it is $\pi_i u_i$ in other periods. In order to derive the terminal cost, it should be pointed out that the value of the electricity remains in the storage after the last period is just $\pi_1$, the lowest price. Thus the terminal cost $-\pi_1 r_{n+1}$ and the total expected cost is:
%The total expected cost is
%for $i^{th}$ period and the following periods is
% \begin{equation}
% J_i= \pi_i (r_i+x_i-r_{i-1})+\sum_{j=i+1}^{n}\pi_j \mathbb{E} (r_j+x_j-r_{j-1}).
% \end{equation}
\begin{equation}
J= \mathbb{E}\left[\sum\nolimits_{i = 1}^n \pi_i(r_i+X_i-r_{i-1}) \right].
\end{equation}
%where we set the reservation in the last period (off peak) as $C$ i.e., fully recharging the storage.
Due to linearity of expectation and cost function, this problem 
satisfies the principle of optimality, which yields our DP formulation. We illustrate the process in Fig. \ref{fig:my_label}.

%\vspace{-0.3cm}
\section{Optimal Storage Sizing}
\label{sec:mainresult}

Based on the DP formulation, we first propose the optimal control policy to utilize the storage system. Then, based on this optimal policy, we solve the optimal storage investment (storage sizing) problem.

\vspace{-0.3cm}
\subsection{Optimal Control Policy}

Note that, in our DP formulation, the difficulties are two folded: the decisions are temporally coupled together, and compared with stationary decision making, the decisions can be made in a dynamic fashion as more information is available. 

Assuming the user has invested in a storage system with capacity $C$. We focused on stationary control policies (control actions are determined at the beginning of the control horizon and are fixed throughout the horizon), and we use the notion of \emph{virtual reservation} to decouple the decisions. Later, we will show how to construct the optimal control policy based on the stationary one.

%In this subsection, we focus on the optimal control policy assuming that we have invested a storage with capacity $C$. We begin with the following definition which is a kind of special control policy.

A control policy with virtual reservation $M_i$ means that the quantity $M_i$ may violate the physical constraints. Hence, at the end of $i^{th}$ period, we project the virtual reservation $M_i$ by the capacity constraints, and reserve at least $\min\{M_i,C\}$ for future use. We first consider a stationary policy, and denote $J_i(M_i|M_{i+1},\cdots,M_n)$ as the expected cost after $i^{th}$ period given reservation $M_{i+1},\cdots,M_n$. We employ backward induction to decide optimal reservations $M_i^*$'s. For simplicity, we define  % if the electricity in the storage is smaller than $\min\{M_i,C\}$ then we charge to $\min\{M_i,C\}$ otherwise we do nothing.
\begin{equation}
    J_i(M_i)= J_i(M_i|M_{i+1}^*,\cdots,M_n^*).
\end{equation}
% We characterize user's control policy in $i^{th}$ period as its reservation for following periods $r_{i+1}^*$:
% \begin{align}
%     r_{i+1}^*=\max\{r_i-x_i,M^{true}_i\},
% \end{align}
% where 
% \begin{align*}
%     M^{true}_i=\min\{C,M^*_i\}.
% \end{align*}
% $M^*_i$ is the virtual optimal reservation of each period. However, the real reservation is limited by the storage capacity and the realized demand. If $M^*_i$ is too big, then we can only charge the storage to $C$. If $M^*_i$ is smaller than the leftover $r_i-x_i$, we just do not charge any more in this period.
%Note that the primal problem is closed-loop (i.e. can make decision according to the state). However, if the user takes a control policy with virtual reservation, it transforms into an opened-loop problem: we just decide a sequence of reservation before the first period which is not related to the exact state in each period. However, we will show that the solution of the opened-loop problem is also an optimal solution of the primal problem.
With such notations, we can show the following lemmas.
\begin{lemma}
If user reserves $M_i$ in $i^{th}$ period, then
\begin{equation}
\frac{\mathrm{d} J_i}{\mathrm{d} M_i} = -\sum\nolimits_{j = i+1}^{n}\pi_j P_j^i(M_i),
\label{equation: lemma1}
\end{equation}
where $P_j^i(M_i)$ denotes the probability that by reserving $M_i,M^*_{i+1},\cdots,M^*_n$ for future use, $j^{th}$ period is the first period that the user need charge its storage system.
\label{lemma:1}
\end{lemma}

\textbf{Remark:} The economic intuition of this lemma is simply to express the total marginal revenue by reserving additional unit of energy in $i^{th}$ period. %This lemma can be explained in an economic perspective. Suppose we reserve $M_i$ in $i^{th}$ period, and the next charge is in $j^{th}$ period. This happens with probability $\text{Pr}\{\text{the first charge after reserve }M_i\text{ is in }\text{$j^{th}$ period}\}$. If we reserve $\epsilon$ kWh more in $i^{th}$ period, then we can charge $\epsilon$ kWh less in $j^{th}$ whose marginal revenue is $\pi_j$. Thus the RHS of (\ref{equation: lemma1}) is the total marginal revenue by total probability rule.

To construct the optimality theorem, we need the following lemma to ensure the uniqueness of our solution.
%One may argue that if there exists a period between $i$ and $j$ whose price is lower than $\pi_j$, then the marginal revenue when the first charge is in $j^{th}$ period will be lower than $\pi_j$ since we can purchase in that period with a lower price. However, this argument is not correct for we are controlling with the optimal reservation. Although we can purchase with a lower price but it is not rational in this case. So the marginal revenue is still $\pi_j$. 
\begin{lemma}
If user reserves $M_i\leq C $ in $i^{th}$ period, then
\begin{equation}
\frac{\mathrm{d}^2J_i}{\mathrm{d} M_i^2}  > 0.
\end{equation}
\label{lemma:2}
\end{lemma}\vspace{-0.5cm}
%Lemma \ref{lemma:2} guarantee the uniqueness of the solution if we let the marginal revenue for continue to reserve equal to the marginal cost. Specifically, if $\pi_{i+1}>\pi_i$ which means the marginal revenue when reserve 0 is greater than the marginal cost, then the optimal reservation exists and is unique. Otherwise it is optimal to reserve 0.
With these two lemmas we can prove that a stationary control policy with virtual reservation is optimal. Specifically, if $\pi_{i+1}>\pi_i$, the optimal reservation exists and is unique. Otherwise, it is optimal not to reserve, and wait for the next period to directly purchase from the grid. In fact, there exists a sequence of virtual reservation $M^*_i$ which is optimal for any capacity $C$. 

To design an algorithm to obtain this sequence, we first identify the boundary condition: the optimal reservation for the last period must be $C$. Then, we assume that the capacity $C$ is sufficiently large such that all the $M_i^*$ will be smaller than $C$. In this case, Lemma \ref{lemma:2} dictates $M_i^*$'s, which yields Algorithm \ref{algorithm: 1}.%  Then it can regard as to reserve $+\infty$ in the last period is optimal. However, for the other periods, the optimal reservation must be finite since the value will become $\pi_n$ which is lower if the reservation will not be used out after the $(n-1)^{th}$ period. We begin with an algorithm to compute the optimal reservation when the capacity is sufficient i.e. greater than the optimal reservation for every periods except the last one. Then we try to show that the output of the algorithm is optimal for any capacity. The algorithm is straightforward from lemma \ref{lemma:2}.

\begin{algorithm}
\caption{Optimal Reservation for $i^{th}$ Period}
\begin{algorithmic}[1] 
\Require $M^*_{i+1},\cdots ,M^*_n$
\Ensure $M^*_i$
\If{$\pi_i\geq \pi_{i+1}$} 
\State \Return 0
\Else
\State solve $M_i$ s.t. $\pi_i = -\frac{\mathrm{d}J_i }{\mathrm{d} M_i}$
\State \Return $M_i$
\EndIf
\end{algorithmic}
\label{algorithm: 1}
\end{algorithm}

%Since the marginal revenue is monotone decreasing, we can simply use binary search to solve the optimal reservation. Then we can run the algorithm \ref{algorithm: 1} from the last period to the second one to compute the optimal reservation for each period when the capacity is sufficient. 

We can solve $M_i^*$ via binary search since the marginal revenue is monotonic decreasing. To see why Algorithm 1 also works for arbitrary $C$, we make the following observation while deciding $M_i^*$'s.

%Up to now we have studied the case when the capacity is sufficient. However, the correctness of the algorithm is not clear for arbitrary capacity yet. Before proceeding further, let us study the following property of algorithm \ref{algorithm: 1} which plays a central role in our analysis.

\begin{lemma}
The optimal reservation $M_i^*$ by Algorithm \ref{algorithm: 1} depends solely on $M_j^*$'s where $M_j^*<M_i^*$.% for a period is not related to an optimal reservation which is greater than or equal to it i.e. suppose the optimal reservation for period $i$ and $j$ are $M^*_i$ and $M^*_j$ respectively. If $M^*_j \geq M^*_i$, then the process to compute $M^*_i$ does not need the exact value of $M^*_j$.
\label{theorem1}
\end{lemma}

That is, for all $M^*_j \geq M^*_i$, we do not need their exact values when deciding $M^*_i$. Hence, combining the physical constraints, we know that all the virtual reservations larger than the capacity $C$ won't affect the decision process of other virtual reservation. This yields the optimality theorem for Algorithm \ref{algorithm: 1}.
%We conclude this subsection by pointing out that the correctness of algorithm \ref{algorithm: 1} when the capacity is arbitrary.
\begin{theorem}
Algorithm \ref{algorithm: 1} dictates the optimal virtual reservation sequence for any capacity $C$.
\label{theorem2}
\end{theorem}
\vspace{-0.2cm}

\subsection{Optimal Sizing for Storage Investment}

It is straightforward to solve the problem from an economic perspective. The marginal cost to invest additional capacity is the amortized cost of storage system, denoted by $\pi_s$. Denote the marginal revenue in $i^{th}$ period by a function in $C$, $MR_i(C)$. Classical economic wisdom immediately illustrates a way to decide optimal $C^*$:
\begin{equation}
\pi_s = \sum\nolimits_{i = 1}^n MR_i(C).
\label{equation: investment}
\end{equation}
All that remains unknown is to decide $MR_i(C)$. In fact, we can use $M_i^*$'s to help us understand  $MR_i(C)$. When $M^*_i\leq C$, more investment won't lead to any more revenue in $i^{th}$ period. More investment leads to more revenue only at those periods whose virtual reservations are larger than the current capacity $C$. This also illustrates why we stick to use virtual reservation throughout the paper. More specifically,
%From the optimal control policy we provided, we can answer how much storage we should buy exactly. The optimal investment can be derived by the same method as employed in the last subsection. Let us first define the marginal revenue for a period.
\begin{equation}
MR_i(C) = 
\begin{cases}
0,\ \ \ \ \ \ \ \ \ \ \ \ \ \ \ \ \ \ \  \text{if }M^*_i\leq C\\ 
-\frac{\mathrm{d} J_i}{\mathrm{d} M_i}\big |_{M_i = C}-\pi_i,\ \text{otherwise}
\end{cases}
\label{eq:MR}
\end{equation}
%This definition is direct: if $M^*_i\leq C$, it makes no change when we continue to invest so the marginal revenue in this period is 0.
%The total marginal revenue for invest $C$ is the sum of each period's marginal revenue. This is correct for theorem \ref{theorem2}: when some of the reservations are restricted by the capacity, the optimal reservations of other periods do not change.
%Since the marginal revenue of each period is monotone decreasing with respect to $C$, the total marginal revenue is also monotone decreasing. We can compute the optimal investment by solving the following equation:
Substituting eq. (\ref{eq:MR}) into eq. (\ref{equation: investment}), we can solve the optimal sizing problem via binary search.

Nonetheless, eq. (\ref{equation: investment}) does not always admit a solution due to a too high amortized cost. We seek to give a sufficient and necessary condition to guarantee a solution to eq. (\ref{equation: investment}). This condition is related to the \emph{local} maximal and \emph{local} minimal prices\footnote{If the electricity price of a period is higher (lower) than the adjacent periods', we call it a local maximal (minimal) price.}.
\begin{corollary}
\label{cor:profittable}

Denote all local maximal prices by $H_1,\cdots,H_m$ and all local minimal prices by $L_1,\cdots, L_m$, then \emph{if and only if}
\begin{equation}
\pi_{s} \le \pi_{\max} = \sum\nolimits_{i = 1}^m(H_i-L_i),
\label{equation: profittable}
\end{equation}
eq. (\ref{equation: investment}) admits a unique solution.
\end{corollary}

\textbf{Remark:} Note that $\pi_{\max}$ is the maximal marginal revenue that one unit of storage capacity can achieve. Hence, it is straightforward to see that when the amortized cost $\pi_s$ is larger than $\pi_{\max}$, then the user would rather not invest in any storage. 

\section{Simulation}
\label{sec:simulation}

\begin{figure}[t]
\centering
\begin{tikzpicture}[xscale=1, yscale =1]
%increasing mode 
\draw[very thick,->] (0,0) -- coordinate (x axis mid) (5,0);
%\node[align=right] at (2,-1) {};
\node at (0.4,2.75) {price};
\draw[very thick,->] (0,0) -- coordinate (y axis mid) (0,3);
%\draw [very thick,black] (0,1.34) --  (1.4,1.34)--(1.4,2.48)--(2.2,2.48)--(2.2,2.08)--(3.4,2.08)--(3.4,2.48)--(3.8,2.48)--(3.8,1.34)--(4.8,1.34);
\draw [draw=none, fill=black, fill opacity=0.2] (0,0) rectangle (1.4,1.34);
\draw [draw=none, fill=black, fill opacity=0.6] (1.4,0) rectangle (2.2,2.48);
\draw [draw=none,fill=black, fill opacity=0.4] (2.2,0) rectangle (3.4,2.08);
\draw [draw=none,fill=black, fill opacity=0.6] (3.4,0) rectangle (3.8,2.48);
\draw [draw=none, fill=black, fill opacity=0.2] (3.8,0) rectangle (4.8,1.34);
\node[anchor=east] at (0,1.34) {\small6.7\textcent};
\node[anchor=east] at (0,2.08) {\small 10.4\textcent};
\node[anchor=east] at (0,2.48) {\small 12.4\textcent};
\node[anchor=north, text=black] at (0,-0.03) {\scriptsize 0AM};
\node[anchor=north, text=black] at (1.4,-0.03) {\scriptsize 7AM};
\node[anchor=north, text=black] at (2.2,-0.03) {\scriptsize 11PM};
\node[anchor=north, text=black] at (3.3,-0.03) {\scriptsize 5PM};
\node[anchor=north, text=black] at (3.9,-0.03) {\scriptsize 7PM};
\node[anchor=north, text=black] at (4.8,-0.03) {\scriptsize 12PM};
%\draw [very thick,blue,dashed] (1.25,0) --  (1.25,1.5);
%\node[anchor=north, text=blue] at (1.25,-0.1) {\scriptsize $\sum_i {(C_i-y_1^i)}^+$};
%\node[anchor=east] at (0,1.5) {\small $\pi'$};
%\node[anchor=east] at (0,3.5) {\small $\pi_h^1$};
%\node[align=left,rotate=90,text=red] at (0.5,3) {\small Demand};
%\node[anchor=south,text=blue] at (1.85,3.5) {\small Supply};
%\filldraw (0.65,1.5) ellipse (1.6pt and 2.3pt);
%\draw[very thick,->] (1.35,2)--(0.72,1.55);
%\node at (1.9,2) {\small equil point};
\end{tikzpicture}
\caption{Sample  multi-peaked ToU pricing scheme.\vspace{-0.2cm}
}
\label{fig: touscheme}
\end{figure}
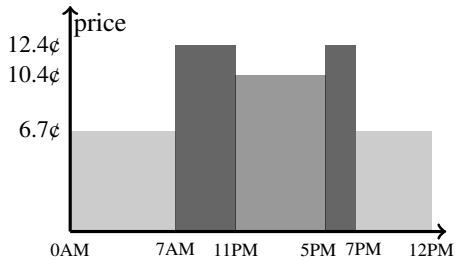

We evaluate the performance of our proposed scheme with real households’ profile (in summer of 2016) from Austin in Pecan Street \cite{Pecan}. We consider a 4-tier multi-peaked ToU pricing scheme from Ontario Energy Board \cite{OEB}, as shown in Fig.  \ref{fig: touscheme}. We assume the amortized cost of storage system is 2 \textcent/kWh. 
%\begin{figure}
%    \centering
%    \includegraphics[width=3in]{touscheme.png}
%    \caption{Four-tier multi-peaked ToU pricing scheme for simulation}
%    \label{fig: touscheme}
%\end{figure}

\begin{figure}[t]
    \centering
    \includegraphics[width=2.7in]{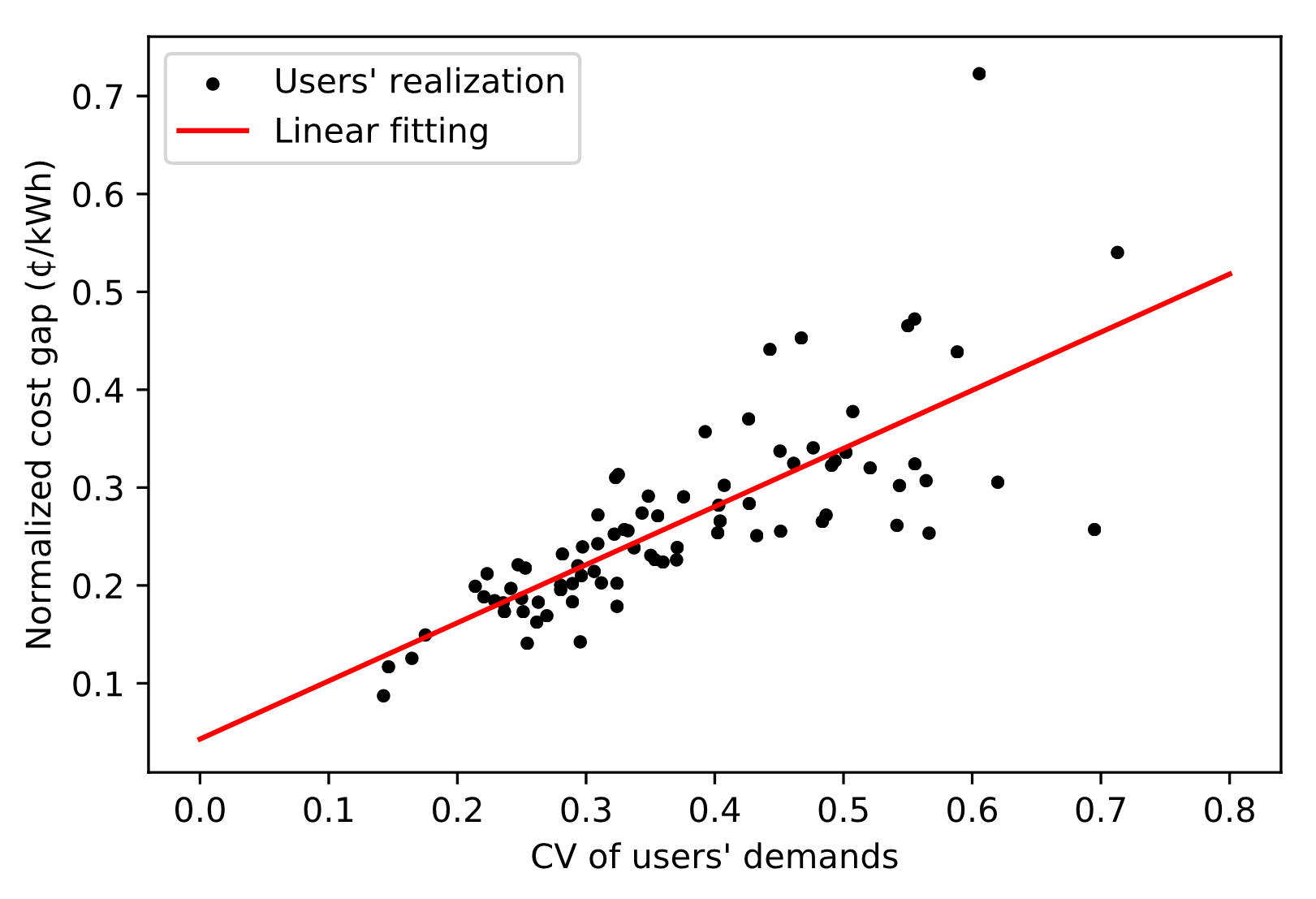}
    \caption{CV v.s. the average cost gap.\vspace{-0.5cm}}
    \label{fig: Deviation}
\end{figure}

\begin{figure}[t]
    \centering
    \includegraphics[width=2.7in]{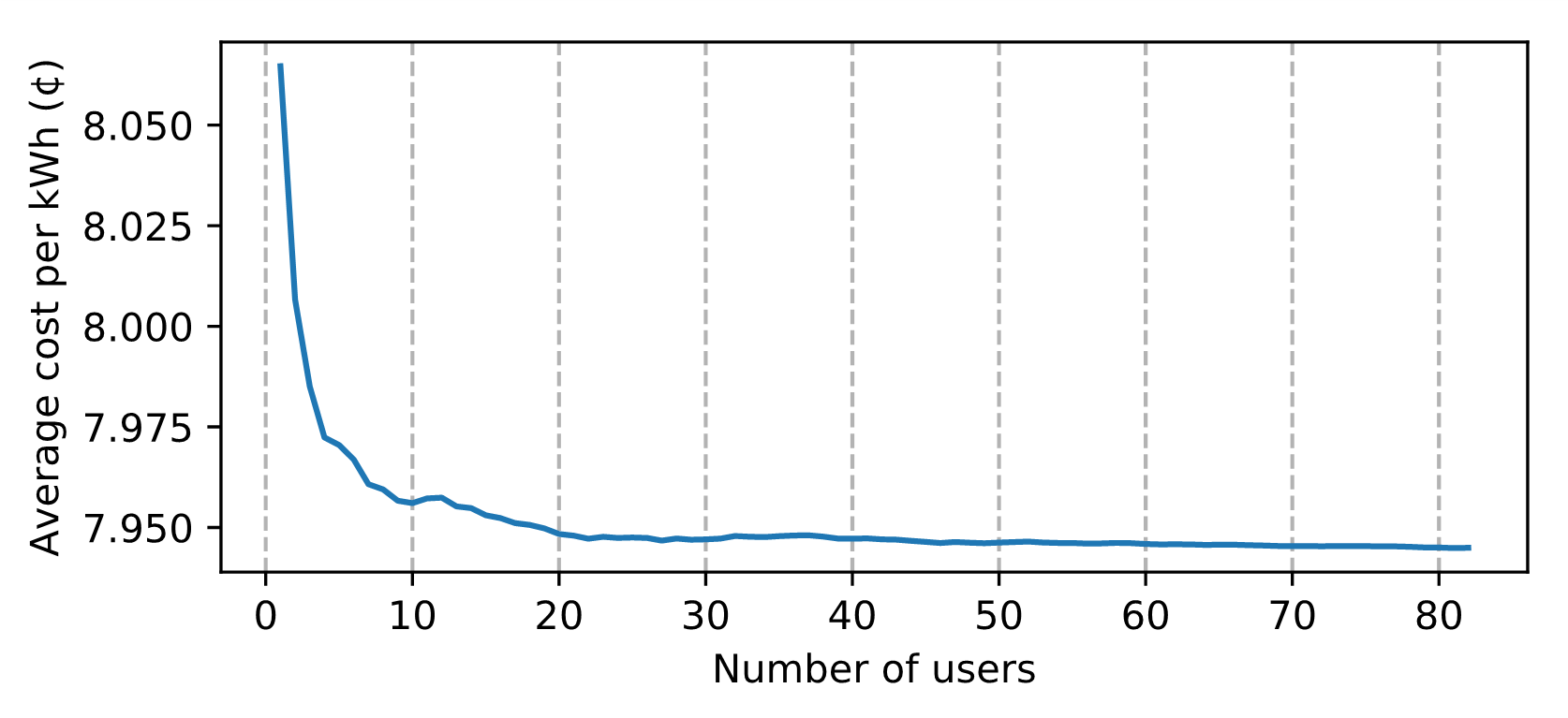}
    \caption{Merge users for more savings.\vspace{-0.5cm}}
    \label{fig: Policycompare}
\end{figure}

Figure \ref{fig: Deviation} examines how the user's cost varies with respect to the user's demand randomness. We use coefficient of variation (CV) to evaluate such randomness. For each user, we calculate the normalized cost gap between the case of random demands and the case of constant demands (the mean value) to demonstrate the cost savings. It is evident that higher CV leads to higher average cost. That is, for each individual user, it is desirable to reduce its randomness in demand.

However, reducing the randomness may not always be feasible for individual user. One possible way is for individuals to cooperate and merge into a single entity for arbitrage. Figure \ref{fig: Policycompare} illustrates that aggregating more users can indeed reduce the average cost. However, such effect is most remarkable till aggregating the first 10 users. This effect diminishes after 20 users' cooperation. This observation can help decide the effective size of aggregator for arbitrage.

\section{Conclusion}
\label{sec:conclusion}
We investigate the optimal control policy and the optimal storage investment facing general multi-tier ToU pricing. We submit that a sequence of virtual reservations can achieve the most electricity bill savings.

There are quite several interesting future directions. We seek to relax our independent demand assumption and provide more theoretical insights. In this case, a control policy with virtual reservation may not be optimal in that historical information will affect our future reservation.

\bibliographystyle{IEEEtran}
\bibliography{IEEEabrv,name}

\section{Appendix}

\subsection{Sketch proof for Lemma \ref{lemma:1}}
Denote the true reservation by $N_i = \min \{M^*_i, C\}$. Then, we know $J_i(M_i) =\sum_{j = i+1}^{n}\pi_j\mathbb{E}[u_j]$, where $u_j$ is the purchase in $j^{th}$ period. More precisely, $\mathbb{E}[u_j]$ can be obtained via the law of total probability:
\begin{equation}
\begin{split}
J_i(M_i) =\sum_{j = i+1}^{n}\pi_j\sum_{\text{all sequence }Q}\mathbb{E}[u_j|Q]\cdot \text{Pr}\{Q\},
%\\=& \sum_{j = i+1}^{n}\pi_j\sum_{\text{all sequence %}Q}\int_{0}^{M_i-N_{i+1}}f_{i+1}(x_{i+1})
%\\&\int_{0}^{M_i-N_{i+1}-x_{i+1}}f_{i+2}(x_{i+2})
%\\&\cdots\int_{M_i-N_{Q_1}-x_{i+1}-\cdots-x_{Q_1-1}}^{+\infty}f_{Q_1}(x_{Q_1})
%\\&\int_{0}^{N_{Q_1}-N_{Q_1+1}}f_{Q_1+1}(x_{Q_1+1})\cdots
%\\&\int_{N_{Q_m}-N_j-x_{Q_m+1}-\cdots-x_{j-1}}^{+\infty}f_k(x_j)(x_{Q_m+1}+\cdots
%\\&+x_j+N_j-N_{Q_m})\mathrm{d}x_j\cdots\mathrm{d}x_{i+1}
\end{split}
\end{equation}
where $Q$ denotes a sequence $i<Q_1<\cdots<Q_m<j$ and we charge at periods $\{Q_1,\cdots,Q_m,j\}$; given $Q$,
$$u_j=N_j-(N_{Q_m}-X_{Q_m+1}-\cdots-X_{j-1}).$$
Taking the derivative with respect to $M_i$ will lead to our conclusion in Lemma \ref{lemma:1}.$\hfill\blacksquare$

\subsection{Sketch proof for Lemma \ref{lemma:2}}

%\textit{\textbf{proof}} Let $N_j = \min \{M^*_j, C\},\forall j>i$ which denote the true reservation of each period.

From lemma \ref{lemma:1} we have 
\begin{equation}
\begin{split}
\frac{\mathrm{d} J_i}{\mathrm{d} M_i} = &-\sum_{j = i+1}^{n}\pi_jP_j^i(M_i)
\end{split}
\end{equation}
%From (\ref{equation: integral1}) and (\ref{equation: integral2}) we can see that the probability can be written in a integral form. 
When computing the second order derivative, the key is to identify that after mathematical manipulation, $\frac{\mathrm{d}^2 J_i }{\mathrm{d} M_i^2}$ has the following form:
%$\forall k\leq j$, denote $D_{jk}$ as the derivative respect to the $M_i$ in the integral bounds of the k-th period in $P_j^i(M_i)$. And we can see that $D_{n,n}$ does not exist so we have:
\begin{equation}
\frac{\mathrm{d}^2 J_i }{\mathrm{d} M_i^2} = \sum\nolimits_{k = i+1}^{n-1}\sum\nolimits_{j = k}^{n}D_{jk}, 
\label{equation: lemma2}
\end{equation}
where 
\begin{equation}
\begin{split}
\sum_{j = k}^{n}D_{jk} = A_k\bigg(\pi_k-\sum_{j = k+1}^{n}\pi_j P_j^k(N_k)\bigg)\geq 0,
\end{split}
\end{equation}
and $A_k$ is a positive coefficient. The last inequality holds due to the optimality of $N_k$. And for period $k=n-1$, this term is strictly positive. Hence, the lemma immediately follows.$\hfill\blacksquare$ 
%We define a variable $A_k$ for convenient which is
%\begin{equation}
%\begin{split}
%A_k &= \int_{0}^{M_i-N_{i+1}}f_{i+1}(x_{i+1})\cdots
%\\&\int_{0}^{M_i-N_{k-1}-x_{i+1}-\cdots-x_{k-2}}f_{k-1}(x_{k-1})\cdot
%\\&f_k(M_i-N_k-x_{i+1}-\cdots-x_{k-1})
%\\&\mathrm{d}x_{k-1}\cdots\mathrm{d}x_{i+1}\geq 0
%\end{split}
%\end{equation}
%Notice that $A_k$ is an integral of some pdf so it cannot be smaller than 0. $\forall k$ we have:
% If $k<j$, we can compute that:
% \begin{equation}
% \begin{split}
% D_{jk} &= -\pi_jA_k\cdot P_j^k(N_k)
% \end{split}
% \end{equation}
% Otherwise if $k = j$, we have
% \begin{equation}
% \begin{split}
% D_{kk} &= \pi_k\cdot A_k
% \end{split}
% \end{equation}
%\begin{equation}
%\begin{split}
%\sum_{j = k}^{n}D_{jk} = A_k\bigg(\pi_k-\sum_{j = k+1}^{n}\pi_j P_j^k(N_k)\bigg)\geq 0
%\end{split}
%\end{equation}

\subsection{Proof of Lemma \ref{theorem1}}
For $j<i$, regardless of exact value of $M^*_j$, it \emph{won't} affect $M^*_i$. Next, we focus on the case when $j>i$.

When $\pi_{i+1}<\pi_i$, it is straightforward to see that we need not reserve anything at $i^{th}$ period, i.e., $M^*_i = 0$. Hence, in this case, regardless of the exact value of $M_j^*$, it won't affect $M_i^*$.% from our algorithm and it cannot be related to $M^*_j$.

Otherwise, $M^*_i$ is the unique solution to 
\begin{equation}
\pi_i = \sum\nolimits_{k=i+1}^{n} \pi_k P_k^i(M_i).
\label{equation: theorem2}
\end{equation}
And it suffices to show that for $M^*_j\geq M^*_i$, all $P_k^i(M_i)$'s are irrelevant to $M_j^*$. By the definition of $P_k^i(M_i)$, it is the probability that $k^{th}$ period is the first period that the user need charge the storage. Due to the fact that $M^*_j\geq M^*_i$,  $P_k^i(M_i)=0, \forall k>j$. For $k<j$, $P_k^i(M_i)$ is not related to $M^*_j$. The only remaining hurdle is when $k=j$. Note that
\begin{equation}
    \sum\nolimits_{k=i+1}^{n} P_k^i(M_i) = 1,
\end{equation}
we know that
\begin{equation}
    P_j^i(M_i) = 1- \sum\nolimits_{k=i+1}^{j-1} P_k^i(M_i),
\end{equation}
which is also irrelevant to $M_j^*$. Thus, we complete the whole proof.$\hfill\blacksquare$ 

\subsection{Proof of Theorem \ref{theorem2}}
We  prove  the  theorem  by  backward  induction. For the last period, the optimal reservation is $C$, which is the boundary of our algorithm, and is optimal. %the output of algorithm 1 must be+∞sinceπnis the lowest price. Thus the conclusion hold.

Suppose the reservation for $i+1^{th}$ period to the last period are all optimal. For $i^{th}$ period, if $M^*_i>C$, then the marginal revenue must be greater than 0, which implies that reserving $C$ is optimal and the conclusion holds. Otherwise, from Theorem \ref{theorem1}, we know that $M^*_i$ is not related to the reservation which is greater than $C$. Thus, it suffices to prove the conclusion when the capacity is sufficiently large. Due to induction hypothesis, we can show that reserving $M^*_i$ is optimal in this case, and also optimal for arbitrary capacity $C$. $\hfill\blacksquare$

\subsection{Proof of Corollary \ref{cor:profittable}}

The maximal profit that unit storage can achieve is
\begin{equation}
    \sum\nolimits_{i = 1}^n \max\{\pi_{i+1}-\pi_i,0\}.
    \label{eq:profit}
\end{equation}

Separating the whole periods by $H_1,\cdots,H_m$ and $L_1,\cdots,L_m$, we can observe that from each local minimal price to local maximal price (rate increasing periods), all the prices between these two periods in (\ref{eq:profit}) can be eliminated. 
For other periods (rate decreasing periods), $\max\{\pi_{i+1}-\pi_i,0\}$ is always zero. These two observations immediately lead to our conclusion.$\hfill\blacksquare$ 
\end{document}